\documentclass[twocolumn]{autart}    % Enable this line and disable the 
\usepackage{graphicx}
\usepackage[utf8]{inputenc}
\usepackage[T1]{fontenc}
\usepackage[english]{babel}
\usepackage{amsmath}
\usepackage{amssymb}
\usepackage{amsfonts}
\usepackage{gensymb}
\usepackage{calc}
\usepackage{empheq}
\usepackage{stackengine}
\usepackage{float}
\usepackage{stmaryrd}
\usepackage{multirow}
\usepackage{wrapfig}
\usepackage{bm}
\usepackage{enumitem}

\newtheorem{remark}{\textbf{Remark}}

\usepackage{mathtools}  
\mathtoolsset{showonlyrefs}

\usepackage{array}
\newcolumntype{L}[1]{>{\raggedright\let\newline\\\arraybackslash\hspace{0pt}}m{#1}}
\newcolumntype{C}[1]{>{\centering\let\newline\\\arraybackslash\hspace{0pt}}m{#1}}
\newcolumntype{R}[1]{>{\raggedleft\let\newline\\\arraybackslash\hspace{0pt}}m{#1}}

\usepackage{epstopdf}

\makeatletter
\renewcommand*\env@matrix[1][*\c@MaxMatrixCols c]{%
	\hskip -\arraycolsep
	\let\@ifnextchar\new@ifnextchar
	\array{#1}}
\makeatother         % Include this line if your 
\begin{document}

\begin{frontmatter}
%\runtitle{Insert a suggested running title}  % Running title for regular 
                                              % papers but only if the title  
                                              % is over 5 words. Running title 
                                              % is not shown in output.

\title{Data-driven Bayesian estimation of Monod kinetics} % Title, preferably not more 
                                                % than 10 words.

\thanks[footnoteinfo]{This work was supported by VINNOVA Competence Center AdBIOPRO, contract [2016-05181]  and by the Swedish Research Council through the research environment NewLEADS (New Directions in Learning Dynamical Systems), contract [2016-06079].}

\author[DCS,AdBIOPRO]{K\'evin Colin}\ead{kcolin@kth.se},    % Add the 
\author[DCS,AdBIOPRO]{H\r{a}kan Hjalmarsson}\ead{hjalmars@kth.se},               % e-mail address 
\author[CETEG,AdBIOPRO]{V\'eronique Chotteau}\ead{veronique.chotteau@biotech.kth.se}  % (ead) as shown

\address[DCS]{Division of Decision and Control Systems, KTH Royal Institute of Technology, Stockholm, Sweden}  % Please supply                                              
\address[CETEG]{Department of Industrial Biotechnology, KTH Royal Institute of Technology, Stockholm, Sweden}             % full addresses
\address[AdBIOPRO]{Competence Centre for Advanced BioProduction by
	Continuous Processing (AdBIOPRO), KTH Royal Institute of Technology, Stockholm, Sweden}       % here.

                             % keyword list or with the 
                                          % help of the Automatica 
                                          % keyword wizard

\begin{abstract}                          % Abstract of not more than 200 words.
 In this paper, we consider the well known problem of non-linear identification of the rates of the reactions involved in cells with  Monod functions. In bioprocesses, generating data is very expensive and long and so it is important to incorporate prior knowledge on the Monod kinetic parameters. Bayesian estimation is an elegant estimation technique which deals with parameter estimation with prior knowledge modeled as probability density functions. However, we might not have an accurate knowledge of the kinetic parameters such as interval bounds, especially for newly developed cell lines. Hence, we consider the case when there is no accurate prior information on the kinetic parameters except qualitative knowledge such that their non-negativity. A log-Gaussian prior distribution is considered for the parameters and the mean and variances of these distribution are tuned using the Expectation Maximization algorithm. The algorithm requires to use Metropolis Hastings within Gibbs sampling which can be computationally expensive. We develop a novel variant of the Metropolis-Hastings within Gibbs sampling sampling scheme in order to accelerate and improve on the hyperparameter tuning. We show that it can give better modeling performances on a relatively large-scale simulation example compared to available methods in the literature.
\end{abstract}

\end{frontmatter}

\section{Introduction}

The last decades have seen an increase in the number of therapeutic treatments based on proteins produced  by mammalian cells. As an example, monoclonal antibodies, produced by Chinese Hamster Ovary cells, are the basis of several medications~\cite{Bre:00,Wal:91} used for the treatment of patients suffering from auto-immune diseases (e.g., Crohn's disease~\cite{Han:06}), from cancers~\cite{Zah:20,Shu:12} and, most recently, from COVID-19~\cite{Llo:21,Tay:21}. In order to meet the increasing demand of proteins, the cells are cultivated in industrial bioreactors for which the temperature, the stirring level and the pH are monitored. The cells are fed with a feed-medium containing some sugars (glucose, galactose, etc) and a range of amino-acids (serine, asparagine, etc).

However, the exploitation cost of such industrial processes is very high and this impacts the price of the therapeutic treatments. That is why there has been a large effort in optimizing bioreactors . Among the several research lines of optimization, the optimization of the feed-medium is one of them. The main idea is to determine the concentrations of the components in the feed-medium such that one of several objectives are met, for example maximization of the productivity of the proteins of interest or minimization of the toxic by-products (e.g., lactate and ammonia which inhibits cell growth). This optimal balance between the concentrations can be determined experimentally by trial-and-error approaches but this implies a large number of experiments which are both expensive and highly time-consuming. Consequently, model-based optimization have been considered in the bioprocessing literature where the optimal feed-medium is determined from a model of the evolution of the concentrations of the metabolites of interest in the bioreactor.

Among all the features that need to be modeled, the kinetics of the cells are one of the most important ones. By cell kinetic estimation, we refer to data-driven modeling of the rates of the reactions involved in the cell metabolism. These models depend on the concentrations of the metabolites such as the cells, the sugars, the amino-acids, the lactate and the ammonium for CHO cells. A detailed-cellular modeling approach is a very challenging task because of the high number of possible reactions between the metabolites. Moreover, the problem suffers from lack of measurements because gathering data of all the metabolites in the bioreactor is an expensive and highly time-consuming task. The macroscopic modeling approach is one way to remedy the aforementioned problems~\cite{YMH:15,CHZM:21}. It consists of describing the kinetics between extracellular metabolites only, leading to considerably simpler models, yet substantial. Even though these models are not directly trained with intracellular measurements, they are still accurately informative with respect to the dynamics of the intracellular metabolites. There are two macroscopic modeling approaches: black-box and grey-box estimation.

In black-box estimation,  the identification of the model solely relies on the data, i.e., no biological principles are used for the estimation of the rates. Several black-box techniques have been used for kinetic estimation such as neural networks~\cite{CBBA:00,GWB:05} and Gaussian process regression~\cite{BSZJD:18,DECBZJ:19}. However such modeling strategies can suffer from both underfitting and overfitting issues and the resulting models can be difficult to interpret biologically. Grey-box modeling approaches consider biochemistry equations as a basis for the estimation of the rates of the biochemical reactions. Several types of grey-box models have been developed in the literature such as Michaelis-Menten mechanistic models~\cite{MiM:1913}. These models take into account the enzyme catalysis effect and consider the sole effect of one metabolite as an activation function. An extension of Michaelis-Menten models was also developed, called Monod models~\cite{Mon:49}. In the Monod model structure, the combined effect of all the metabolites affecting the reaction rate is modeled by a product of rational functions, each of them depending on only one metabolite concentration~\cite{HHHJC:17}. Moreover each rational function is not restricted to be an activation function since three other types of kinetics can be considered: inhibition, double-component and neutral effect. These rational functions depend also on some kinetic constants which are the parameters to be estimated using data. Therefore, the grey-box identification problem consists in estimating the kinetic constants by solving a nonlinear least-squares optimization with measurements of metabolite concentrations and rates. However this problem is non-convex which implies that local minimum issues may happen. For Michaelis-Menten models (i.e., activation with one metabolite), there are several linearizations techniques via transformations (Eadie–Hofstee, Hanes–Wolf, Lineweaver–Burk, inverse Eadie–Hofstee, see Section 3.1 of~\cite{Tou:16} for a brief review of these transformations and further references). However, it becomes hard to scale up these methods for Monod models with several metabolites. A linearization technique has been developed in~\cite{Wan:19} for double-component structures with one metabolite. Since Monod model estimation is a rational function identification problem, one can think of transforming the non-convex problem  into a linear one by multiplying the identification criterion by the denominator. In~\cite{Wan:19} an algorithm providing unbiased estimates has been developed. However, if we want to consider reactions rates of higher order in order to model more interactions between the metabolites, the number of linear parameters to be identified increases rapidly with respect to the number of metabolites and so this approach becomes computationally unrealistic for real-life kinetic models.

The second line of research for solving the non-convex least-squares problem is to initialize the non-convex algorithms accurately. In other words, by using an intermediate modeling approach, we compute a first estimate of the parameters which is close to the global optimum of the non-convex least-squares objective function. Several intermediate methods have been developed such as graphical approaches (which only works for a small number of metabolites), Bayesian estimation approaches~\cite{Her:19} or grey-box approaches based on  Gaussian Processes as in~\cite{WRJCH:20}. The approach in the latter is to keep the product of the individual effects as modeling basis but each modulation function is identified as a Gaussian process with a covariance function specifically designed for the identification of Monod effects. This kernel works well for large amount of data which are widely distributed in the concentration space. However, most of the time, there is not a lot of data in bioprocessing applications since it is expensive and highly time-consuming to do experiments. In~\cite{Col:22}, we developed a better-tailored kernel function for Monod kinetic effects in order to solve the problem of poorly identifiable problems and we showed that it works on a small toy example with six modulation functions. Nevertheless, when scaling up the estimation problem to larger metabolic network with, e.g., twelve metabolites, the method does not yield accurate results anymore and becomes computationally expensive.

In this paper, we propose to use the Bayesian estimation framework for the estimation of Monod kinetic parameters. Our work differs from the one in~\cite{Her:19} where accurate prior information in the form of uncertainty intervals was assumed to be available for the to be estimated kinetic parameters. Contrary to~\cite{Her:19} we propose in this paper to consider the scenario where there is no available accurate prior information on the kinetic parameters and we will just use basic prior information such as the non-negativity of the kinetic parameters and the fact they can vary by several orders of magnitude. In other words, we want to use the Bayesian estimation method by using the data only. After defining a suitable class of parametric prior distributions for the kinetic parameters, we estimate the corresponding hyperparameters with the Empirical Bayes method~\cite{Cas:85}, i.e., we compute the hyperparameters maximizing the log-marginal likelihood function. 

As it is often the case for nonlinear parametric Bayesian  estimation, we need to introduce some latent variables in order to simplify the maximization procedure of the Empirical Bayes method. The latent variables are estimated together with the hyperparameters by using the Expectation Maximization (EM) algorithm. However, some Markov chain Monte Carlo (MCMC) sampling methods are needed in order to approximate the intractable integral of the E-step of the EM algorithm. The idea is to sample from the posterior distribution with a fixed value for the hyperparameters and use these samples in order to approximate the integral.  The sampling method must be able to circumvent two issues: $(a)$ the number of kinetic parameters may be large especially for large metabolic networks with a large number of metabolites and $(b)$ the normalization distribution present in the posterior distribution is intractable. The Metropolis-Hastings within Gibbs sampling~\cite{Gew:01} (also called the single-component Metropolis-Hastings sampling) is a sampling method which tackles both aforementioned issues problems. The candidate samples of the Metropolis-Hastings step can get often rejected, which can give a slow convergence of estimation.  Therefore, in this article, we develop a novel variant of the Metropolis-Hastings within Gibbs sampling in order to accelerate the hyperparameter tuning. The main idea is to repeat the Metropolis-Hastings step until a candidate sample is accepted. We call this sampling \textit{enforced} Metropolis-Hastings within Gibbs sampling. We illustrate the good performances of the novel sampling schemes compared to the classical Metropolis-Hastings within Gibbs sampling and  Gaussian process methods~\cite{WRJCH:20,Col:22} in a relatively large numerical example. 

To sum up, the main contributions of this work are $(i)$ the proposal of a novel prior suited for the Bayesian estimation of the kinetic parameters in Monod models when very poor knowledge is available beforehand, $(ii)$  a novel variant of the Metropolis Hastings within Gibbs sampling method for better and faster convergence of the hyperparameter tuning and $(iii)$ the demonstration of the performance of the method compared to other approaches on a numerical example of relatively large complexity (12 metabolites) not previously considered in the literature.

% OR

%\begin{figure}
%\begin{center}
%\epsfig{file=jcaesar,width=7cm}
%\caption{Gaius Julius Caesar, 100--44 B.C.}
%\label{fig1}
%\end{center}
%\end{figure}

\section{Notations}

The set of real-valued vectors $x$ of dimension $n$ will be denoted $\mathbb{R}^n$ and the set of real-valued non-negative vectors $x$ of dimension $n$ will be denoted $\mathbb{R}_+^n$. We denote by $||x||_2$ is the $\mathcal{L}_2$ norm of any vector $x\in\mathbb{R}^n$.

\section{Macroscopic kinetic modeling with Monod functions}

\subsection{Monod kinetics}

In this paper, we consider the modeling of the rate of a macroscopic reaction  involved in some cells cultivated in a bioreactor. We will denote by $m$ the total number of extracellular metabolites (substrates or products) in the bioreactor which can influence the kinetics of the this reaction. The concentration of the $i$-th metabolite will be denoted by $c_i$. 

\begin{figure}[H]
	\centering
	\includegraphics[width = \linewidth]{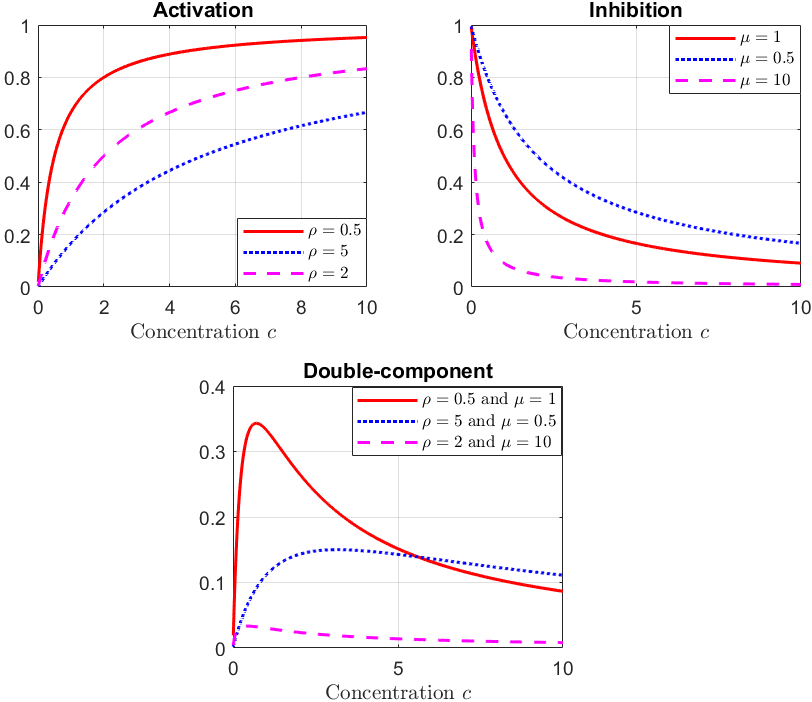}
	\caption{Different activation (top left), inhibition (top right) and double-component (bottom) functions.}
	\label{fig:monod}
\end{figure}

Denote by $w$ the rate of this reaction. We will assume that the rate $w$ is of the Monod-type~\cite{BaK:89,Les:03,HWR:05,HHHJC:17,HHMFC:19}. In that case, $w$ is expressed as the product of $m$ modulation functions $h_i$ 
\begin{equation}\label{eq:w}
	w(c) = \alpha \prod_{i=1}^m h_i(c_i)
\end{equation}where $\alpha$ is the maximal rate constant. Each function $h_i$ describes the individual effect of the $i$-th metabolite on the kinetics. Moreover, each $h_i$ can only have 4 different parametric rational expressions:
\begin{equation*}
	\begin{array}{rlrl}
		(i) \ \  &  \dfrac{c_i}{c_i + \rho_i} \ \ \ \ \ \ \ \ \ & (ii)  \ \ &\dfrac{1}{1 + \mu_i c_i} \\
		(iii) \ \ & \dfrac{c_i}{c_i + \rho_i} \dfrac{1}{1 + \mu_i c_i} \ \ \ \ \ \ \ \ \ \ &(iv) \ \ & 1
	\end{array}
\end{equation*}where $\rho_{i}$ and $\mu_{i}$ are respectively half saturation and half inhibition constants (kinetic parameters). The expression $(i)$ corresponds to the activation effect: the metabolite accelerates the reaction when its concentration increases. A metabolite with an activation effect cannot increase infinitely the rate and so a saturation effect must be considered. This is modeled by the fact that the activation function converges to $1$ when the concentration $c_i$ goes to infinity. Contrary to the activation effect, some metabolites can decelerate the reaction when they get more concentrated: it is the inhibition effect, modeled by $(ii)$. As an example, too much lactate is harmful for Chinese Hamster Ovary cells~\cite{Buch:18}, i.e., lactate has an inhibition effect on the growth rate. Some metabolites can combine both effects and this is called the double-component effect. The corresponding modulation function has the expression $(iii)$ which is the multiplication of an activation function by an inhibition function. Finally, some metabolites may have no effect on the kinetics of $w$: it is the neutral effect modeled by $(iv)$. In Figure~\ref{fig:monod}, we plot different activation, inhibition and double-component functions.

\begin{remark}\label{rem:dc} The activation, the inhibition and the neutral functions are particular cases of the double component one with either or both parameters being equal to 0.
\end{remark}

In the next paragraph, we detail the assumptions on the measurements and the considered identification approach for $w$.

\subsection{Data and identification}

We assume that we have some observations of $w$ at $N$ different time instants $t=1,\cdots,N$ of the form $y(t) = w(c(t)) + e(t)$ where $y$ is the measured rate and $e$ is a  zero-mean white  Gaussian noise of variance $\sigma_{e}^2$. We will also assume that we have \textit{noiseless} measurements of the concentrations $c_i$ for the $m$ metabolites at the same time instants as $w$. This may seem like a strong assumption but is commonly used in the field as the noisy case is very difficult. Finally, we consider that the concentrations $c_i$ are independent from the noise $e$.

Since the type of kinetic effect of each metabolite is not known beforehand and by using the fact that the activation $(i)$, the inhibition $(ii)$ and the neutral effect $(iv)$ are particular cases of the double component $(iii)$ (see Remark~\ref{rem:dc}), we consider a double component function for all modulation functions $h_i$. Consequently, we have to estimate the following parameters: $\alpha$, $\{ \rho_i\}_{i=1}^m$ and $\{ \mu_i\}_{i=1}^m$. We will denote by $\bm{\theta}$ the vector collecting the kinetic parameters $\{ \rho_i\}_{i=1}^m$ and $\{ \mu_i\}_{i=1}^m$. There are therefore $2m+1$ parameters to be identified ($2m$ in $\bm{\theta}$ and the maximal rate constant $\alpha$).

Based on the data, we can build the non-linear least-squares estimates $\hat{\bm{\theta}}$ and $\hat{\alpha}$ for the kinetic parameters as shown below
{\begin{align}
	&\{\hat{\bm{\theta}},\hat{\alpha}\}  = \underset{\substack{\bm{\theta}\ \ge\ 0\\\alpha \ \ge \  0}}{\text{ arg min }} \sum_{t=1}^N(y(t) - {w}(c(t),\bm{\theta},\alpha))^2\label{eq:nls_opti}\\
	  &{w}(c(t),\bm{\theta},\alpha) = \alpha \prod_{i=1}^m h(c_i(t),\rho_i,\mu_i) \ \ \  \label{eq:nls_w}\\
  &h(c_i(t),\rho_i,\mu_i) = \dfrac{c_i(t)}{c_i(t) + \rho_i} \dfrac{1}{1 + \mu_i c_i(t)}\label{eq:hi}
\end{align}}However, the model ${w}(c(t),\bm{\theta},\alpha)$ is non-convex with respect to $\{ \rho_i\}_{i=1}^m$ and $\{ \mu_i\}_{i=1}^m$ which makes the optimization problem non-convex. Some local minimum issues are then expected with classical non-convex optimization algorithms such as the ones developed for non-linear least-squares optimization, e.g. the Levenberg-Marquardt algorithm~\cite{Mar:63,Mor:78} or Newton-type algorithms specialized for rational function estimation~\cite{Dim:91}.

This can be solved by initializing these algorithms close to the global optimum. In the next paragraph, we detail the approach considered in~\cite{WRJCH:20} in order to compute a good initial estimate of the parameter vector. As mentioned in the introduction, a first intermediate modeling method is multilinear Gaussian process regression where each modulation function $h_i$ is modeled as a Gaussian process~\cite{WRJCH:20,Col:22}. However, as will be illustrated in Section~\ref{sec:numerical_toy_example}, this approach can fail when the number $m$ of metabolites to be considered increases due to the high modeling flexibility inherent with the Gaussian process method. In~\cite{Her:19}, the Bayesian framework was leveraged for the estimation of Monod parameters but accurate prior knowledge in the form of bounds on the kinetic parameters are required. In this paper, we consider the scenario where no prior information is known on the kinetic parameters and the identification solely relies on the gathered data. Indeed, it can be common to not have prior knowledge on all the kinetic parameters, especially for new cell lines. Therefore, in this paper, we propose to draw on the Bayesian estimation framework for the identification of the Monod parameters $\bm{\theta}$ and $\alpha$ by  solely using the data and qualitative knowledge on the kinetic parameter such that their non-negativity.

\begin{remark}
    Even though the proposed approach is developed for the worst-case scenario when no quantitative information is available  on the kinetic parameters such as bounds, it can also be adapted to this case.
\end{remark}

\section{Bayesian estimation - a general presentation}

In this section, we give a brief general explanation of the Bayesian method dedicated to parameter estimation problems. In the next section, we show how to apply this method for the particular problem of Monod kinetic identification.

\subsection{Prior and posterior distributions}\label{sec:Bayes_general_prior}

In Bayesian estimation, we are interested in estimating some parameters collected in a vector $\bm{\Theta}$ by using some random observations collected in a vector $\bm{Y}$. The main difference of the Bayesian framework as compared to the frequentist approach with maximum likelihood techniques) is that the parameter vector $\bm{\Theta}$ is considered to be a random variable.

However, we may have some prior knowledge on this parameter vector which can be useful to be exploited to get a more reliable estimate of $\bm{\Theta}$. Bayesian estimation allows to include this prior knowledge in the estimation process in a stochastic manner. Indeed, in the Bayesian framework, the parameter vector $\bm{\Theta}$ is assumed to be a random variable following some probability distribution function called \textit{prior} distribution denoted by $\pi(\bm{\Theta})$. This probability density function corresponds to our belief (or prior knowledge) on the parameter $\bm{\Theta}$ before any observation of data.

The prior distribution can be parametrized and the parameters are called hyperparameters. We can adapt the prior density function accordingly to our prior knowledge by adequately tuning these hyperparameters. As an example, if we consider a Gaussian distribution for the prior distribution, then the variance and the mean are the hyperparameters which can be tuned. We will denote by $\bm{\eta}$ the vector collecting all the hyperparameters and by $\pi(\bm{\Theta};\bm{\eta})$ the parametrized prior density function. 

The observed output data $\bm{Y}$ carry some information about the unknown parameter vector $\bm{\Theta}$. This information can be combined with the prior information by defining the posterior distribution $p(\bm{\Theta}|\bm{Y} ;\bm{\eta})$ for $\bm{\Theta}$ given the observed data $\bm{Y}$. In other words, the posterior distribution can be seen as the update of our prior belief on $\bm{\Theta}$ after the observation of the data $\bm{Y}$. The posterior distribution is given by the Bayes rule:
\begin{equation}\label{eq:posterior}
	p(\bm{\Theta}|\bm{Y};\bm{\eta}) = \dfrac{p(\bm{Y} | \bm{\Theta}) \pi(\bm{\Theta};\bm{\eta})}{p(\bm{Y})}
\end{equation}where $p(\bm{Y} | \bm{\Theta};\bm{\eta})$ is the likelihood distribution and $p(\bm{Y}) = \int_{\bm{\Theta}} p(\bm{Y} | \bm{\Theta};\bm{\eta})\pi(\bm{\Theta};\bm{\eta})d\bm{\Theta} $ is the normalization distribution which is independent from $\bm{\Theta}$. Therefore, the posterior is proportional to $p(\bm{Y} | \bm{\Theta};\bm{\eta}) \pi(\bm{\Theta};\bm{\eta})$ and the normalization constant is here to guarantee that the posterior distribution of the random variable $\bm{\Theta}$ has a total probability mass equal to 1.

Now that we have defined the posterior distribution, we will explain in the next paragraph how to construct a Bayesian estimator. 

\subsection{Loss function, risk and Bayes estimator}

In order to compute an appropriate estimator $\hat{\bm{\Theta}}(\bm{Y})$, we first need to define the loss function $L(\bm{\Theta},\hat{\bm{\Theta}}(\bm{Y}))$. This function gives the loss of choosing the estimator $\hat{\bm{\Theta}}(\bm{Y})$ while the true value of the parameter is $\bm{\Theta}$. The most commonly used loss functions is the quadratic loss $L(\bm{\Theta},\hat{\bm{\Theta}}(\bm{Y})) = ||\bm{\Theta}-\hat{\bm{\Theta}}(\bm{Y})||^2_2$.

From the loss function $L(\bm{\Theta},\hat{\bm{\Theta}}(\bm{Y}))$, we can define the  posterior expected loss (or risk) of choosing $\hat{\bm{\Theta}}(\bm{Y})$ instead of the true value after observation of the data $\bm{Y}$ as follows
\begin{align}\label{eq:bayes_risk}
	R(\hat{\bm{\Theta}}(\bm{Y}) | \bm{Y}) &= \int_{\bm{\Theta}}L(\bm{\Theta},\hat{\bm{\Theta}}(\bm{Y}))p(\bm{\Theta}|\bm{Y};\bm{\eta}) d\bm{\Theta}
\end{align}We define the optimal estimator $\hat{\bm{\Theta}}^\star(\bm{Y})$ minimizing the risk, i.e., 
\begin{equation}\label{eq:bayes_estimator}
	\hat{\bm{\Theta}}^\star(\bm{Y}) = \underset{\hat{\bm{\Theta}}(\bm{Y})}{\text{ arg min }}  R(\hat{\bm{\Theta}}(\bm{Y}) | \bm{Y})
\end{equation}The latter is called a Bayes estimator.

In order to apply this method for the estimation of $\bm{\Theta}$, we need to choose the hyperparameter vector $\bm{\eta}$ of the prior distribution. As aforementioned in Section~\ref{sec:Bayes_general_prior}, these hyperparameters must be tuned w.r.t. our prior knowledge of the parameter $\bm{\Theta}$. In the case where no knowledge is available, it is possible to tune $\bm{\eta}$ by using the data $\bm{Y}$. The most common approach is the Empirical Bayes~\cite{Cas:85} and it is the one we will consider in this paper for the Monod kinetic estimation. We give a brief recap of this method in the next paragraph.

\subsection{Empirical Bayes and Expectation Maximization algorithm for hyperparameter tuning}

The principle of the Empirical Bayes method~\cite{Cas:85} is to choose the hyperparameter vector $\hat{\bm{\eta}}$ which maximizes the log (marginal) likelihood:
\begin{align}
	\hat{\bm{\eta}} &= \underset{\bm{\eta}}{\text{ arg max }}  \ell(\bm{\eta}) \ \ \text{ with } \ \ \ell(\bm{\eta}) = \log(p(\bm{Y} |\bm{\eta}))
\end{align} For linear estimation problem with additive Gaussian noise and with a Gaussian prior function, the latter can be computed exactly. However, as is very often the case in nonlinear Bayesian estimation, there is no available closed-form expression of $\ell(\bm{\eta})$ and so it is not possible to solve the maximization analytically. 

%By using the chain rule $p(\bm{y},\bm{\xi}|\bm{\eta}) = p(\bm{\xi}|\bm{y},\bm{\eta})p(\bm{y}|\bm{\eta})$, the log-likelihood  $\log(p(\bm{y},\bm{\xi}|\bm{\eta}))$ can be expressed as
%\begin{equation}
%    \log(p(\bm{y},\bm{\xi}|\bm{\eta})) = \log(p(\bm{\xi}|\bm{y},\bm{\eta})) +  \log(p(\bm{y}|\bm{\eta}))
%\end{equation}and so $ =  \log(p(\bm{y}|\bm{\eta}))$ is given by
%\begin{equation}\label{eq:log_marg_like_for_EM}
%    \ell(\bm{\eta}) = \log(p(\bm{\xi},\bm{y}|\bm{\eta}))) -\log(p(\bm{\xi}|\bm{y},\bm{\eta})) 
%\end{equation}

The classical solution circumventing this issue is to introduce some latent variables\footnote{Latent variables are random variables which are unobserved and which will be estimated together with $\bm{\eta}$.} (collected in a vector $\bm{\xi}$) which will simplify the computation of $\ell(\bm{\eta})$. The Expectation Maximization algorithm proposed in~\cite{DLR:77_EM} is an iterative procedure which, at each iteration, builds a lower bound for the log-marginal likelihood $\ell(\bm{\eta})$ and then maximizes this lower bound, which in turn maximizes $\ell(\bm{\eta})$. At the $j$-th iteration of the EM algorithm and by denoting by $\hat{\bm{\eta}}^{(j)}$ the value of the hyperparameter obtained at the end of iteration $j$, a lower bound $Q(\bm{\eta})$ for $\ell(\bm{\eta})$ is given by
\begin{align}\label{eq:lower_bound_Q}	Q(\bm{\eta}|\hat{\bm{\eta}}^{(j-1)}) = \int \log(p(\bm{\xi},\bm{Y}|\bm{\eta}))p(\bm{\xi}|\bm{Y},\hat{\bm{\eta}}^{(j-1)})d\bm{\xi} \ \ 
\end{align}i.e., it is obtained by taking the expected value of $\log(p(\bm{\xi}|\bm{Y},\bm{\eta}))$ with respect to the current conditional distribution of $\bm{\xi}$ given the data $\bm{Y}$ and the estimate $\hat{\bm{\eta}}^{(j-1)}$ computed at the end of iteration $j-1$. The task of computing~\eqref{eq:lower_bound_Q} is called the E-step. Then, a new value  $\hat{\bm{\eta}}^{(j)}$ for the hyperparameter can be computed by maximizing $Q(\bm{\eta}|\hat{\bm{\eta}}^{(j-1)})$ and this maximization is called the Q-step. To sum-up, starting from an initial value $\bm{\eta}^{(0)}$ for the hyperparameter vector $\bm{\eta}$, the EM algorithm produces a sequence of $\hat{\bm{\eta}}^{(j)}$ obtained  by performing iteratively the following two steps:
{\begin{align*}
		\text{E-step: }   & Q(\bm{\eta}|\hat{\bm{\eta}}^{(j-1)}) = \int \log(p(\bm{\xi},\bm{y}|\bm{\eta}))p(\bm{\xi}|\bm{y},\hat{\bm{\eta}}^{(j-1)})d\bm{\xi}\\
		\text{Q-step: }  &   \hat{\bm{\eta}}^{(j)} = \underset{\bm{\eta}}{\text{ arg max }} Q(\bm{\eta}|\hat{\bm{\eta}}^{(j-1)})
\end{align*}}It is known that the sequence $\{\hat{\bm{\eta}}^{(j)}\}_{j\ge 0}$ converges to a local maximum of $\ell(\bm{\eta})$ where convergence to saddle points are occasional~\cite{DLR:77_EM}.

For most problems, the calculation of $Q(\bm{\eta}|\hat{\bm{\eta}}^{(j-1)})$ in~\eqref{eq:lower_bound_Q} is impossible due to the presence of intractable integrals. Markov Chain Monte Carlo approaches~\cite{Gil:95,Chi:01} are computing methods which can leverage this issue by approximating an integral calculation such as $Q(\bm{\eta}|\hat{\bm{\eta}}^{(j-1)})$. The idea is to sample $L$ different latent variable vectors $\hat{\bm{\xi}}^{(j,k)}$ $(k=1,\cdots,L)$ from $p(\bm{\xi}|\bm{y},\hat{\bm{\eta}}^{(j-1)})$ and to use them to approximate $Q(\bm{\eta}|\hat{\bm{\eta}}^{(j-1)})$ as follows
\begin{equation}\label{eq:Qk_approx}
	Q(\bm{\eta}|\hat{\bm{\eta}}^{(j-1)}) \approx \dfrac{1}{L}\sum_{k=1}^L \log(p(\hat{\bm{\xi}}^{(j,k)},\bm{y}|\hat{\bm{\eta}}^{(j-1)}))
\end{equation}We will see in the next section how to adapt the Bayesian estimation method with the EM algorithm for the Monod estimation.

\section{Bayesian estimation applied to Monod kinetic modeling}

In this section, we explain how to apply the Bayes estimation on our problem of Monod kinetic modeling, i.e., the problem of the estimation of the kinetic parameter vector $\bm{\theta}$ and the maximal constant rate $\alpha$. Hence,
\begin{equation}
\bm{\Theta} = \begin{pmatrix}
    \bm{\theta}\\ \alpha
\end{pmatrix}
\end{equation}

As explained in the last section, we need to define $(i)$ a family of parametrized prior distributions, $(ii)$ a loss function, $(iii)$ a computation procedure of the corresponding Bayes estimator and $(iv)$ the latent variables for the hyperparameter estimation. In the next three paragraphs, we elaborate on the choices considered in this paper.

\subsection{Choice of the prior}

As aforementioned, the prior distribution should be designed based on the known information about the kinetic parameters. In this paper, we consider the worst-case when we have little prior knowledge about the kinetic parameters. We will only consider the two following facts which always hold for any kinetic parameter: 
\begin{itemize}
    \item they are non-negative.
    \item they can range over several orders of magnitude.
\end{itemize}

There are several classical distributions satisfying the non-negativity requirement: truncated Gaussian distribution, log-Gaussian distribution, Beta distribution, Gamma distribution, etc. In~\cite{Her:19}, a Gamma distribution was considered. However, the kinetic parameters can vary between several orders of magnitude (usually between $10^{-3}$ and $10^2$) and a Gamma distribution does not handle this well. In this paper, we propose to use the logarithmic Gaussian distribution,  However, we will only consider a log-distributed prior for the half saturation constants $\rho_i$ and the half saturation constants $\mu_i$ as given below
{\begin{align}
	\rho_i& \sim \dfrac{
1}{\rho_i\sqrt{2\pi}\sigma_{\rho_i}}\exp\left(-\dfrac{(\log(\rho_i) - \beta_{\rho_i})^2}{2\sigma_{\rho_i}^2}\right) \\
	\mu_i&  \sim \dfrac{
1}{\mu_i\sqrt{2\pi}\sigma_{\mu_i}}\exp\left(-\dfrac{(\log(\mu_i) - \beta_{\mu_i})^2}{2\sigma_{\mu_i}^2}\right)
\end{align}}where $\beta_{\rho_i}$ and $\beta_{\mu_i}$ are the mean of the log-distribution of $\rho_i$ and $\mu_i$ respectively and where  $\sigma_{\rho_i}$ and $\sigma_{\mu_i}$ are the standard deviation of the log-distribution of $\rho_i$ and $\mu_i$ respectively. Those means and variances are hyperparameters that we need to estimate with the data.

For the maximal rate constant $\alpha$, we will explain later how to estimate it accurately without the need of defining any prior distribution.

\subsection{Likelihood function}

With the zero-mean white Gaussian assumption of the noise $e$, the likelihood $p(\bm{y} | \bm{\theta},\alpha)$ is given by
{\begin{equation}\label{eq:likelihood}
	 \dfrac{1}{(2\pi\sigma_e^2)^{\frac{N}{2}}}\exp\left(-\dfrac{\sum_{t=1}^N(y(t) - {w}(c(t),\bm{\theta},\alpha))^2}{2\sigma_e^2}\right)
\end{equation}}where $w(c(t),\bm{\theta},\alpha)$ is defined in~\eqref{eq:nls_w}. In this case, the parameter vector $\hat{\bm{\theta}}$ and gain $\hat{\alpha}$ solution of~\eqref{eq:nls_opti}-\eqref{eq:nls_w} maximizes the likelihood distribution $p(\bm{y} | \bm{\theta},\alpha)$, i.e., the non-linear least-squares estimator is also the maximum likelihood estimator and as already noted this method suffers  from the non-convexity of $\bar{w}(c(t),\bm{\theta},\alpha)$. 

\subsection{Choice of the loss function and computation procedure of the Bayes estimator}

For the loss function, we will consider the classical quadratic loss:
\begin{equation}
	L(\bm{\Theta},\hat{\bm{\Theta}}(\bm{y})) = ||\bm{\Theta}-\hat{\bm{\Theta}}(\bm{y})||^2_2 
\end{equation} The corresponding risk function $ R(\hat{\bm{\Theta}}(\bm{y}) | \bm{y})$ is the so-called mean square error (MSE) and the corresponding Bayesian estimator $\hat{\bm{\Theta}}^\star_{MSE}(\bm{y})$ is called the minimum mean square error (MMSE) estimator. For the quadratic loss, there is no need to use optimization algorithms to solve~\eqref{eq:bayes_estimator}. Indeed, $\hat{\bm{\Theta}}^\star_{MSE}(\bm{y})$ is equal to the mean of the posterior distribution $p( \bm{\Theta} |\bm{y},\bm{\eta})$ (See, e.g.,~\cite{Zho:15} for the proof).

\subsection{Adaptation of the EM algorithm for the hyperparameter estimation linked to kinetic parameters}

In order to apply the EM algorithm in the problem of Monod kinetic estimation, we need to define the latent variables $\bm{\xi}$ and to specify the computation of both E- and Q-steps. As proposed in~\cite{McK:07}, we will estimate the hyperparameters together with the parameters related to the prior distributions, i.e., $\bm{\xi} = \bm{\Theta}$.

The calculation of  $Q(\bm{\eta}|\hat{\bm{\eta}}^{(j-1)})$ in~\eqref{eq:lower_bound_Q} is however intractable in our case. Therefore, we have to consider the approximation in~\eqref{eq:Qk_approx} by using some samples obtained from the posterior $p(\bm{\xi}|\bm{y},\hat{\bm{\eta}}^{(j-1)}) = p(\bm{\Theta}|\bm{y},\hat{\bm{\eta}}^{(j-1)})$ given by
\begin{equation}
	p(\bm{\Theta}|\bm{y},\hat{\bm{\eta}}^{(j-1)}) = \dfrac{p(\bm{y} | \bm{\Theta};\hat{\bm{\eta}}^{(j-1)}) \pi(\bm{\Theta};\hat{\bm{\eta}}^{(j-1)})}{p(\bm{y})}
\end{equation}Since the hyperparameters we choose are the mean and the variance of the log-Gaussian distributions, the Q-step becomes easy to realize. Let us explain it for the computation of the hyperparameters $\hat{\beta}_{\rho_1}^{(j)}$ and $\hat{\sigma}_{\rho_1}^{(j)}$ during the Q-step at the $j$-th iteration of the EM-algorithm. Consider $L$ samples $\hat{\rho}_1^{(j,k)}$ ($k=1,\cdots,L$) obtained during the E-step during the $j$-th iteration. The update of the hyperparameters $\hat{\beta}_{\rho_1}^{(j)}$ and $\hat{\sigma}_{\rho_1}^{(j)}$ is done as follows
\begin{align}
	\hat{\beta}_{\rho_1}^{(j)} &= \dfrac{1}{L}\sum_{k=1}^{L} \log\left(\hat{\rho}_1^{(j,k)}\right)\\
	\hat{\sigma}_{\rho_1}^{(j)} &=\sqrt{ \dfrac{1}{L}\sum_{k=1}^{L} \left(\log\left(\hat{\rho}_1^{(j,k)}\right)- \hat{\beta}_{\rho_1}^{(j)}\right)^2}
\end{align}The computation of the mean and the variance for the other parameters is similar. However, getting samples from the posterior distribution $	p(\bm{\Theta}|\bm{y},\hat{\bm{\eta}}^{(j-1)})$ is not straightforward for two reasons:
\begin{enumerate}[label=$(\alph*)$]
	\item the number of variables in $\bm{\theta}$ can be large, especially for metabolic networks with a large amount\footnote{In genome scale metabolic networks REF, one can expect more than 1000 metabolites.} of metabolites.
	\item the estimation of the normalization distribution $p(\bm{y})$ is most of the time computationally expensive.
\end{enumerate}

In the next paragraph, we present the proposed sampling strategy in order to  tackle both problems $(a)$ and $(b)$.

\subsection{Metropolis-Hastings within Gibbs sampling and the proposed variation}

The idea is to combine two existing sampling methods which can individually solve  one issue each. We present both sampling techniques in the next two paragraphs and we explain later how to combine them in order to solve $(a)$ and $(b)$ together. 

\subsubsection{Tackling (a) with Gibbs sampling} 

Gibbs sampling is a technique which tackles the sampling of large-scale random vector~\cite{Gem:84,Rou:97_gibbs,Gel:00}. The idea is to iteratively sample one random variable at a time while keeping the other ones equal to their previous sampled values. When this variable has been sampled, the next one is sampled and one repeatedly cycles through all the variables. 

\begin{figure}[H]
	\centering
	\includegraphics[width = \linewidth]{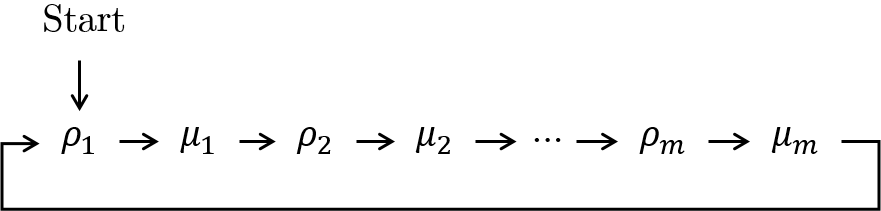}
	\caption{Chosen order for Gibbs sampling applied to Monod estimation.}
	\label{fig:gibbs}
\end{figure}

In our application, we will sample as follows: we start with the first modulation function ($h_1$) and we sample the half saturation parameter $\rho_1$. Then, we sample the half inhibition constant $\mu_1$. After that, we switch to the second modulation function ($h_2$) and we sample successively $\rho_2$ and $\mu_2$. We go on like this until the last modulation function $h_m$. When we have sampled the parameters $\rho_m$ and $\mu_m$, we repeat the sampling of both parameters $\rho_1$ and $\mu_1$ and we go on like this repeatedly. This is illustrated in Figure~\ref{fig:gibbs}. Note that we do not include the maximal rate constant $\alpha$ in the Gibbs loop. As will be explained later in Section~\ref{sec:max_rate_constant}, this parameter will be adapted every time a kinetic parameter in $\bm{\theta}$ is sampled.

\begin{remark}
	The convergence of the Markov chain is independent from the sampling order of the variables.
\end{remark}

The individual sampling of each variable is done by considering its respective conditional posterior distribution, i.e., we sample a given parameter from the posterior distribution $p(\bm{\theta}|\bm{y};\hat{\bm{\eta}}^{(j-1)})$ for which all the other variables are set equal to their previously sampled value. Let us clarify this procedure with an example with $m=2$ metabolites. Denote by $\hat{\rho}_1^{(j,k)}$, $\hat{\mu}_1^{(j,k)}$, $\hat{\rho}_2^{(j,k)}$ and $\hat{\mu}_2^{(j,k)}$ the  sampled values of $\rho_1$, $\mu_1$, $\rho_2$ and $\mu_2$ respectively during the $k$-th loop of the Gibbs sampling and the $j$-th iteration of the EM algorithm. By following the sampling order depicted in Figure~\ref{fig:gibbs}, these samples are obtained as follows{\begin{align*}
		\hat{\rho}_1^{(j,k)} &\sim p(\rho_1 | \bm{y}, \hat{\mu}_1^{(j,k-1)},\hat{\rho}_2^{(j,k-1)}, \hat{\mu}_2^{(j,k-1)};\hat{\bm{\eta}}^{(j-1)})\\
		\hat{\mu}_1^{(j,k)} &\sim p(\mu_1 | \bm{y}, \hat{\rho}_1^{(j,k)},\hat{\rho}_2^{(j,k-1)}, \hat{\mu}_2^{(j,k-1)};\hat{\bm{\eta}}^{(j-1)})\\
		\hat{\rho}_2^{(j,k)} &\sim p(\rho_2 | \bm{y}, \hat{\rho}_1^{(j,k)}, \hat{\mu}_1^{(j,k)}, \hat{\mu}_2^{(j,k-1)};\hat{\bm{\eta}}^{(j-1)})\\
		\hat{\mu}_2^{(j,k)} &\sim p(\mu_2 | \bm{y}, \hat{\rho}_1^{(j,k)},  \hat{\mu}_1^{(j,k)}, \hat{\rho}_2^{(j,k)};\hat{\bm{\eta}}^{(j-1)})
\end{align*}} Note that the first iteration of the Gibbs sampling ($k=1$) requires an initial parameter vector $\hat{\bm{\theta}}^{(0)}$. We will argue later on the choice we have considered in Section~\ref{sec:initial}.

The sampled parameter vectors $\{\hat{\bm{\theta}}^{(j,k)}\}_{k\in\mathbb{N}}$ form a Markov chain which converges to a stationary distribution equal to $p(\bm{\theta}|\bm{y};\hat{\bm{\eta}}^{(j-1)})$.  In practice stationarity is assumed to hold after a sufficient number of repetitions (called burn in period). Moreover, for each individual parameter in $\bm{\theta}$, the samples are uncorrelated.

However, every conditional distribution of the posterior distribution $p(\bm{\theta}|\bm{y};\hat{\bm{\eta}}^{(j-1)})$ depends on the normalization distribution $p(\bm{y})$ which is computationally intractable to get. In the next paragraph, we detail the sampling method commonly used to deal with this problem: Metropolis-Hastings sampling.

\subsubsection{Tackling (b) with Metropolis-Hastings sampling}

Metropolis-Hastings algorithm is a sampling method which approximates the sampling from a desired distribution by the sampling of a proposal distribution~\cite{Hit:03_MH,Met:53,Has:70}. However, it requires the knowledge of a probability distribution which is proportional to the desired one. We satisfy this requirement for the considered problem since every conditional distribution of $p(\bm{\theta}|\bm{y};\hat{\bm{\eta}}^{(j-1)})$ is proportional to the intractable normalization factor $p(\bm{y})$. 

In order to simplify the presentation, we will explain the principle of the Metropolis-Hastings algorithm for the kinetic parameter $\rho_1$ sampled from the conditional posterior distribution $p(\rho_1|\bm{y})$ where all the other parameters are kept constant (we abusively dropped the other parameters in the notation). Denote by $p(\rho_1,\bm{y})$ the joint density function of $\rho_1$ and $\bm{y}$ defined as $p(\rho_1,\bm{y}) = p(\rho_1|\bm{y})\times p(\bm{y})$. Finally, consider an initial value $\hat{\rho}_1^{(0)}$ and a proposal distribution $g(a|b)$ which generates a value $a$ given a previous sampled value $b$. Then, at each iteration $l$ of the Metropolis-Hastings algorithm, we perform the three following steps
\begin{itemize}
	\item Step 1: Sample a candidate value $\hat{\rho}_1'$ from the proposal distribution $g(\hat{\rho}_1'|\hat{\rho}_1^{(l-1)})$.
	\item Step 2: Compute the ratio \begin{equation}
		\gamma = \dfrac{f(\hat{\rho}_1'|\bm{y})g(\hat{\rho}_1^{(l-1)}|\hat{\rho}_1')}{f(\hat{\rho}_1^{(l-1)})|\bm{y})g(\hat{\rho}_1'|\hat{\rho}_1^{(l-1)})}  
	\end{equation}
	\item Step 3: Generate a sample $u$ from an uniform distribution within the range $[0,1]$. Then, 
	\begin{itemize}
		\item If $\alpha \ge u$, we accept the candidate sample $\hat{\rho}_1'$ as a sample from $p(\rho_1|\bm{y})$ and we set $\hat{\rho}_1^{(l)} = \hat{\rho}_1'$.
		\item If $\alpha < u$, we reject the candidate sample $\rho_1'$ as a sample from $p(\rho_1|\bm{y})$ and we set $\hat{\rho}_1^{(l)} = \rho_1^{(l-1)}$.
	\end{itemize}
\end{itemize}The main idea behind this algorithm is to move in the sample space towards high probability regions.

\subsubsection{Metropolis-Hastings within Gibbs sampling}

Both problems $(a)$ and $(b)$ can be simultaneously solved by using the classical Metropolis-Hastings within Gibbs sampling (C-MHWGS)~\cite{Gew:01}. The main idea of the algorithm is as follows: the Gibbs sampling is used in order to sample each parameter successively. The sampling of each parameter is performed by considering only one iteration of the Metropolis-Hastings algorithm, i.e., we sample only one candidate parameter and we perform the acceptance/rejection test of Step~3. 

\begin{figure}
	\centering
	\includegraphics[width = \linewidth]{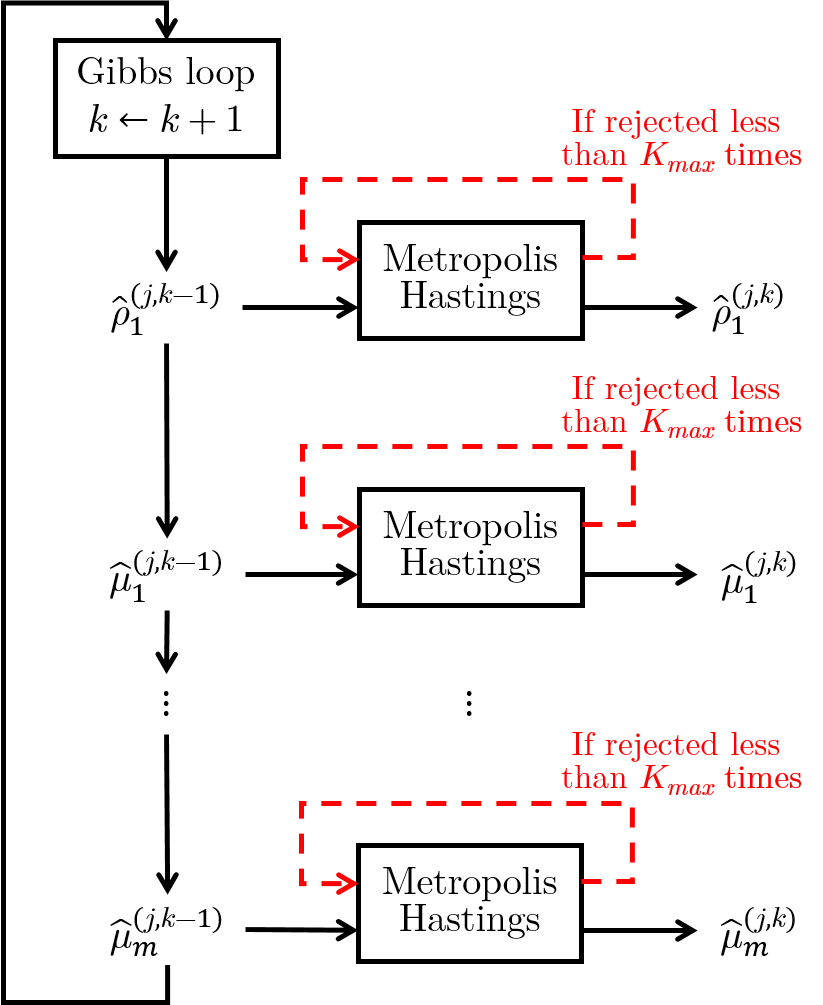}
	\caption{Scheme of the E-MHWGS sampling}
	\label{fig:emhwgs}
\end{figure}

Let us explain more in details for the sampling of the parameter $\rho_1$ during the iteration (or loop) number $k$ of the Gibbs sampling and the $j$-th iteration of the EM-algorithm. The parameter $\hat{\rho}_1^{(j,k)}$ is obtained by using one iteration of the Metropolis-Hastings for which the initial value is $\hat{\rho}_1^{(j,k-1)}$, i.e., the sample obtained during the previous loop of the Gibbs sampling. 
Note that, depending on the result of the acceptance/rejection test of Step 3, $\hat{\rho}_1^{(j,k)}$ may be equal to $\hat{\rho}_1^{(j,k-1)}$. 

Because only one candidate sample is considered for the Metropolis-Hastings algorithm in the C-MHWGS, this algorithm may not explore fast enough the parameter space because of sample rejections possibly leading to slow convergence. In the next paragraph, we propose a variation of this sampling scheme. 

\subsubsection{Proposed variation of the classical Metropolis-Hastings within Gibbs sampling}

In the case of rejection at Step 3 of the Metropolis-Hastings algorithm, the idea is to re-run Steps 1 and 2  several times until one candidate sample gets accepted. In order to avoid too many attempts of the Metropolis-Hastings algorithm, we consider a maximum number $K_{max}$ of  attempts for acceptance of a candidate sample.  We will call this \textit{enforced} Metropolis-Hastings within Gibbs sampling (E-MHWGS). It is illustrated in Figure~\ref{fig:emhwgs}. As will be illustrated in Section~\ref{sec:numerical_toy_example} in a numerical example, this scheme can accelerate the tuning of the hyperparameters and give better fit performances. 

We still need to address two issues in order to realize the sampling: the choice of the proposal distribution used for each Metropolis-Hastings iteration and the estimation of the maximal rate $\alpha$.

\subsubsection{Chosen proposal function}

For the proposal distribution of each parameter in $\bm{\theta}$,  we choose a log-Gaussian distribution $g^{(j,k)}(.|.)$ which is updated at each iteration $(j,k)$. This distribution is taken centered with respect to the logarithmic of the previously sampled value of the corresponding parameter. The standard deviation  of this distribution will be equal to the standard deviation computed at the previous EM iteration of the prior log-Gaussian distribution of the parameter to be sampled. We add an additive small perturbation term $\delta$ to the standard deviation to keep an active exploration of the parameter space. Let us clarify with the example of the kinetic parameter $\rho_1$ sampled during iteration $(j,k)$. The  proposal distribution $g^{(j,k)}(\rho_1'|\hat{\rho}_1^{(j,k-1)})$ is a log-Gaussian with a mean of $\log(\hat{\rho}_1^{(j,k-1)})$ and a standard deviation equal to $\sigma_{\mu_1}^{(j-1)} + \delta$. 

\subsubsection{Sampling of the maximal rate constant $\alpha$}\label{sec:max_rate_constant}

The maximal rate constant $\alpha$ is the only kinetic parameter which appears linearly in the Monod function expression $w(c(t),\bm{\theta},\alpha)$. We will use this fact to our advantage for its estimation. The idea is to compute the maximal rate constant $\alpha$ maximizing the likelihood every time a kinetic parameter in $\bm{\theta}$ is sampled. Let us explain with the sampling of the parameter $\rho_1$ during iteration $(j,k)$. Assume we have a candidate sample ${\rho}_1'$. The optimal value $\hat{\alpha}$ which maximizes the conditional likelihood  where all the kinetic parameters in $\bm{\theta}$ are replaced by their last sample is given by
{\small\begin{align*}
		\hat{\alpha}&= \dfrac{\sum_{t=1}^N y(t)\bar{w}(c(t))}{\sum_{t=1}^N  \bar{w}^2(c(t))}\\
		\bar{w}(c(t)) &= h(c_1(t),\hat{\rho}_1',\hat{\mu}_1^{(j,k-1)})\prod_{i=1}^m h(c_i(t),\hat{\rho}_i^{(j,k-1)},\hat{\mu}_i^{(j,k-1)})
\end{align*}}where  $h(c_i(t),\rho_i,\mu_i)$ is defined by~\eqref{eq:hi} for any $\rho_i$ and $\mu_i$. This result is obtained by determining the parameter $\alpha$ which nullifies the gradient of the conditional likelihood with respect to $\alpha$. The derivation of $\hat{\alpha}$ when the other kinetic parameters are sampled is similar.

To sum up, every time a kinetic parameter in $\bm{\theta}$ is sampled as a candidate, we automatically set $\alpha$ equal to the value maximizing the conditional likelihood. In other words, we sample two candidate parameters during each Metropolis-Hastings iteration: a kinetic parameter and the corresponding maximal rate constant. Then, we perform the acceptance/rejection test of the Metropolis-Hasting algorithm with both $\alpha$ and the kinetic parameter equal to their candidate values.  The motivation of sampling $\alpha$ this way is to $(i)$ avoid to add this parameter in the Gibbs loop so we can reduce the computation time and $(ii)$ it always maximizes the conditional likelihood.

Regarding the EM algorithm, we have to solve two other issues: the computation of the estimate of the noise variance $\sigma_e^2$ at each iteration of the EM algorithm and the initial values for both the hyperparameters $\bm{\eta}$ and the parameters $\bm{\theta}$ and $\alpha$.

\subsection{Estimate of the noise variance $\sigma_e^2$}

Each iteration $j$ of the EM algorithm depends on the estimate $\hat{\sigma}_e^{(j-1)}$ of the standard deviation of the white Gaussian noise $e$ obtained at iteration $j-1$. Our choice will be as follows: at the end of the E-step, we first compute the posterior mean of each kinetic parameter from the samples, i.e., we compute 
\begin{align}
   \hat{\rho}_i^{(j)} &=  \dfrac{1}{L}\sum_{k=1}^{L} \hat{\rho}_i^{(j,k)}\ \text{ and }\ \hat{\mu}_i^{(j)} &= \dfrac{1}{L}\sum_{k=1}^{L} \hat{\mu}_i^{(j,k)}
\end{align}Secondly, we compute the corresponding maximal rate constant $\hat{\alpha}^{(j)}$ similarly as in Section~\ref{sec:max_rate_constant} but by replacing the kinetic parameters with their respective posterior mean estimates $\hat{\rho}_i^{(j)}$ and $\hat{\mu}_i^{(j)}$. Finally, we estimate the standard deviation $\hat{\sigma}_e^{(j)}$ of the modeling error with these posterior estimates, i.e.,  $\hat{\sigma}_e^{(j)}$ is derived as follows
\begin{align}\label{eq:estimate_sigma_e}
    \hat{\sigma}_e^{(j)} &= \sqrt{\dfrac{1}{N}\sum_{t=1}^{N} \left(y(t)- {w}(c(t),\hat{\bm{\theta}}^{(j)},\hat{\alpha}^{(j)})\right)^2}
\end{align}

\subsection{Choice of the initial value $\hat{\bm{\eta}}^{(0)}$ of the hyperparameters and the initial value $\hat{\bm{\theta}}^{(0)}$ of the parameters}\label{sec:initial}

In~\cite{Col:22}, we have assumed that in most cases the kinetic parameters will be between $0.01$ and $100$, so we could set the log-mean and log-variance according to this assumption. In this paper, we instead initialize the hyperparameters in $\bm{\eta}$ by studying the sensitivity of the parameters with respect to the data. Let us first consider the hyperparameters $\beta_{\rho_i}$ and $\sigma_{\rho_i}$ linked to the half saturation constants $\rho_i$. Consider any activation function $ c_i/(c_i+\rho_i)$ and some time instants $t=1,\cdots,N$ for which we have data of the concentration $c_i(t)$. Denote $\underline{c}_{i}$ and $\overline{c}_{i}$ the minimal and maximal value among the data  $\{c_i(t)\}_{t=1}^N$. The activation function is the most sensitive to $\rho_i$ on the interval $[0.1\underline{c}_{i},10\overline{c}_i]$. Therefore, we can choose an initial value $\hat{\beta}_{\rho_i}^{(0)}$ for the mean $\beta_{\rho_i}$ equal to $(\log(0.1\underline{c}_{i})+\log(10\overline{c}_i))/2$ and an initial value $\hat{\sigma}_{\rho_i}^{(0)}$ for the the standard deviation $\sigma_{\rho_i}$ such that $P(\log(0.1\underline{c}_{i}) < \rho_i < \log(10\overline{c}_i)) = 0.95$, i.e., $\hat{\sigma}_{\rho_i}^{(0)} = (\log(10\overline{c}_i)-\log(0.1\underline{c}_{i}))/3$. 

For the half inhibition parameters $\mu_i$, let us observe that we can write the inhibition function $1/(1+\mu_ic_i(t))$ as $1/c_i(t)/(1/c_i(t)+\mu_i)$. It is thus equal to an activation function except that $c_i(t)$ is replaced by $1/c_i(t)$. Therefore, we will choose $\hat{\beta}_{\mu_i}^{(0)}$ and $\hat{\sigma}_{\mu_i}^{(0)}$  similarly as in the activation function case detailed above. The minimal and maximal value of $\{1/c_i(t)\}_{t=1}^N$ are $1/\overline{c}_i$ and $1/\underline{c}_i$ respectively. Therefore, we choose $\hat{\beta}_{\mu_i}^{(0)} = (\log(0.1/\overline{c}_i)+\log(10/\underline{c}_i))/2$ and $\hat{\sigma}_{\mu_i}^{(0)} = (\log(10/\underline{c}_i)-\log(0.1/\overline{c}_i))/3$.

We also need an initial estimate $\hat{\bm{\theta}}^{(0)}$ of the kinetic parameters $\bm{\theta}$ for the E-step of the first iteration of the EM algorithm. For that, we will consider the following initial estimates
\begin{align}
    \hat{\rho}_i^{(0)} &= \text{exp}(\hat{\beta}_{\rho_i}^{(0)}) \ \ \ \text{and} \ \ \ \hat{\mu}_i^{(0)} = \text{exp}(\hat{\beta}_{\mu_i}^{(0)})
\end{align}i.e., the kinetic parameters corresponding to the initial value of the mean of the log-Gaussian priors distributions. For the initial value $\hat{\alpha}^{(0)}$ of the maximal rate constant, we do as in Section~\ref{sec:max_rate_constant} except that $\bm{\theta}$ is replaced by $\hat{\bm{\theta}}^{(0)}$. Finally, for the initial value of the standard deviation $\sigma_e$ of the noise, we will compute it as in~\eqref{eq:estimate_sigma_e} with $j-1$ replaced by $0$, i.e., using the initial values described above.

\section{The problem of parameter identifiability in Monod structures}\label{sec:identif}

Before considering an example in order to compare the proposed technique with the ones in the literature, it is important to address the problem of parameter identifiability in Monod structures. Parameter identifiability is the property of uniquely inferring the true values of the parameters ($\bm{\theta}_0$ and $\alpha_0$ in our case) of a model structure from an infinite number of observations from it. Mathematically, for Monod functions, we would have identifiability if and only if the following property was satisfied for all $(\bm{\theta}^\top,\alpha)^\top\in\mathbb{R}^{2m+1}_+$
\begin{equation}
  {w}(c,\bm{\theta},\alpha) = {w}(c,\bm{\theta}_0,\alpha_0) \ \ \  \forall c\in\mathbb{R}_+^m \implies \left\{ \begin{array}{rcl}
  \bm{\theta} &=& \bm{\theta}_0\\
  \alpha &=& \alpha_0
  \end{array}\right. 
\end{equation}
Unfortunately, this property is not guaranteed for Monod model structures when double component kinetics are present. To illustrate this, let us consider a Monod double structure model structure with $m=1$ metabolite\footnote{We have dropped the index $1$ since only one metabolite is considered.}
\begin{equation}
  {w}(c,\bm{\theta}_0,\alpha_0) = \alpha_0\dfrac{c}{c + \rho_0} \dfrac{1}{1 + \mu_0 c}
\end{equation}and let us observe that we can rewrite it as follows
\begin{equation}
	{w}(c,\bm{\theta}_0,\alpha_0) = \dfrac{\alpha_0}{\rho_0\mu_0}\dfrac{1}{\dfrac{c}{\rho_0} + 1} \dfrac{c}{\dfrac{1}{\mu_0} +  c}
\end{equation}which gives us a new double-component structure with a maximal rate constant equal to ${\alpha_0}/({\rho_0\mu_0})$, a half saturation parameter of $1/\mu_0$ and a half inhibition parameter of $1/\rho_0$. Therefore, they are always two possibilities for the true  half inhibition and half saturation  parameters for double-component kinetics. Hence we have in total $2^{n_{dc,0}}$ different true parameter vectors where $n_{dc,0}$ is the number of double-component functions in the true kinetics. However, we note that lack of parameter identifiability will not cause problems from an optimization perspective as the different true parameters form a set of isolated points in the parameter space. Furthermore, we remind the reader that the key objective is to estimate the kinetics, i.e., the function $w$, which is not affected by the ambiguity in the parameters.

\section{Example with $m=12$ metabolites}\label{sec:numerical_toy_example}

In the previous section, we have argued all the choices for the Bayesian estimation of Monod functions. We will now show its estimation performances on a relatively large scale example. We will compare the porposed scheme E-MHWGS with the classical MHWGS (C-MHWGS) and the proposed sampling scheme E-MHWGS of this paper. We will also add both Gaussian process regression methods derived in~\cite{WRJCH:20} (we will call it GP-2019) and in~\cite{Col:22} (we will call it GP-2022) in the comparative study. 

\subsection{Kinetics}

 Let $w$ be a macroscopic rate  with $m=12$ metabolites with various Monod kinetic effects as described in Table~\ref{tab:kin_example}.
{\begin{table}
	\centering
 \caption{Kinetic types and parameters considered for the $m =12$ modulation functions.}
	\begin{tabular}{c|c|c|c}
		$h_i$ & Kinetic effect & $\rho_i$ & $\mu_i$ \\ \hline
		$h_1$  & Activation &  $0.610$  & $-$  \\ 
		$h_2$  & Inhibition & $-$ & $30.370$ \\ 
		$h_3$  & Double-component & $0.790$ & $1.550$ \\ 
		$h_4$  & Neutral & $-$ & $-$ \\ 
		$h_5$  & Double-component & $0.490$ & $0.280$ \\ 
		$h_6$  & Neutral & $-$ & $-$ \\ 
		$h_7$  & Activation & $0.370$ & $-$ \\ 
		$h_8$  & Neutral & $-$ & $-$ \\ 
		$h_9$  & Activation & $0.760$ & $-$ \\ 
		$h_{10}$  & Inhibition & $-$ & $0.012$\\
            $h_{11}$  & Neutral & $-$ & $-$\\
            $h_{12}$  & Neutral & $-$ & $-$\\
         \end{tabular}
	
	\label{tab:kin_example}
\end{table}}
For the maximal rate constant, we choose $\alpha_0 = 1000$. For the model structure, recall that we consider all modulation functions $h_i$ as double-components (see~\eqref{eq:nls_w}) and there are $2m+1 = 25$ kinetic parameters to be identified. The white noise variance $\sigma_{e}^2$ is taken equal to $0.0001$. 

\subsection{Parameters of the EM and sampling algorithms}

The four modeling methods are all based on the EM algorithm and the Gibbs sampler. Consequently, we will choose the same parameters for a fair comparison. They are given in Table~\ref{tab:param_EM_Gibbs}. In order to accelerate the EM algorithm for the four methods, we will only consider a burn-in period for the first EM iteration, i.e.,  $L_{bi} \ne 0$ at $j = 1$ and $L_{bi} = 0$ for all $j > 1$.
\begin{table}
	\centering
 \caption{Parameters of the EM algorithm and the various sampling methods chosen for the simulations.}
	\begin{tabular}{C{4.3cm}|C{3cm}}
		Parameter & Value\\ \hline
		Number of EM iterations $M$ & 100 \\\hline
		Burn-in $L_{bi}$ for Gibbs sampling &  500 for the first EM iteration, then 0\\\hline
		Number of Gibbs samples $L$ after burn-in  & 100\\\hline
		Number $K_{max}$ of maximal trials for Metropolis-Hastings sampling (only E-MHWGS) & 50\\ \hline
  Perturbation term $\delta$ for the proposal distribution (only C-MHWGS and E-MHWGS) & 0.02
	\end{tabular}
	
	\label{tab:param_EM_Gibbs}
\end{table}

\subsection{Data and computer performances}

We perform 100 Monte Carlo simulations with different concentration data and noise sequences. For each estimation scenario, we choose $N=20$ data where all concentration data $c_i(t)$ are randomly chosen from a positive multivariate truncated Gaussian distribution with a mean vector whole all values are 0.4 and a covariance matrix whose minimal and maximal eigenvalues are $ 1.34\times 10^{-5}$ and $ 1.40\times 10^{-1}$, i.e., some metabolite concentrations are relatively highly correlated. The concentration data belong in the interval $[0,1]$ with a probability of $99\%$.

  The parameters for the EM algorithm and Gibbs sampler are the ones in Table~\ref{tab:param_EM_Gibbs}. The simulations are run on \textsc{Matlab} R2021b with a computer equipped with the processor Intel(R) Core(TM) i5-8365U CPU, 1.60GHz, 4 cores and with 16.0GB of RAM.

\subsection{Criteria of comparison}

 We will base the comparison of the four methods on three different criteria : 
 \begin{itemize}
     \item the fit in $\%$ of the macroscopic rate model defined by
     {\scriptsize\begin{equation*}
          100 \left(1 - \dfrac{\sqrt{\sum_{t=1}^N\left(y(t) - {w}\left(c(t),\hat{\bm{\theta}}^{(M)},\hat{\alpha}^{(M)}\right)\right)^2}}{\sqrt{\sum_{t=1}^N(y(t)-\bar{y})^2}}\right)
     \end{equation*}}where $\bar{y}$ the average of the $N$ output data $y(t)$.
     \item the fit\footnote{Because some modulation functions are neutral effects and so constant, we do not subtract the average of the true modulation function data $h_i({c}_i(t))$ in the denominator terms to avoid a division by 0.} in $\%$  of each modulation function $h_i$ 
     {\scriptsize\begin{equation*}
         100 \left(1 - \dfrac{\sqrt{\sum_{t=1}^N\left(h_i({c}_i(t)) - \lambda_i {h}\left({c}_i(t),\hat{\rho}_i^{(M)},\hat{\mu}_i^{(M)}\right)\right)^2}}{\sqrt{\sum_{t=1}^Nh_i({c}_i(t))^2}}\right) 
     \end{equation*}}with ${h}({c}_i(t),\hat{\rho}_i^{(M)},\hat{\mu}_i^{(M)})$  as defined in~\eqref{eq:hi} and $\lambda_i$ a proportional constant tuned such that the error $\sum_{t=1}^N(h_i({c}_i(t)) - \bar{h}_i({c}_i(t),\hat{\rho}_i^{(M)},\hat{\mu}_i^{(M)}))^2$ is minimized. The reason for this re-scaling is due to the identifiability issue mentioned in Section~\ref{sec:identif} as there might be two possibilities for $h_i$.
     \item the computation time in seconds.
 \end{itemize}

\subsection{Results}

\begin{figure}
    \centering
    \includegraphics[width=1\linewidth]{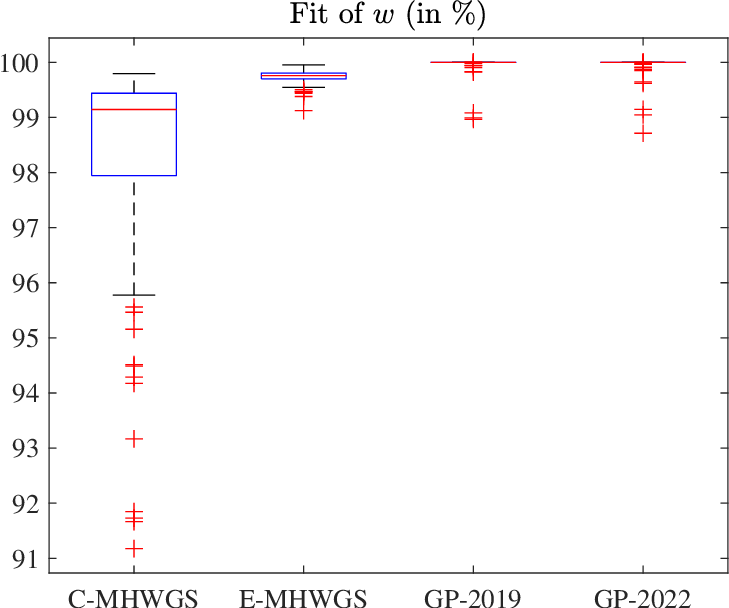}
    \caption{Box plots of the fit obtained with C-MHWGS, E-MHWGS, GP-2019 and GP-2022 with the 100 Monte Carlo simulations.}
    \label{fig:best_fit_boxplot}
\end{figure}

\begin{figure}
    \centering
    \includegraphics[width=1\linewidth]{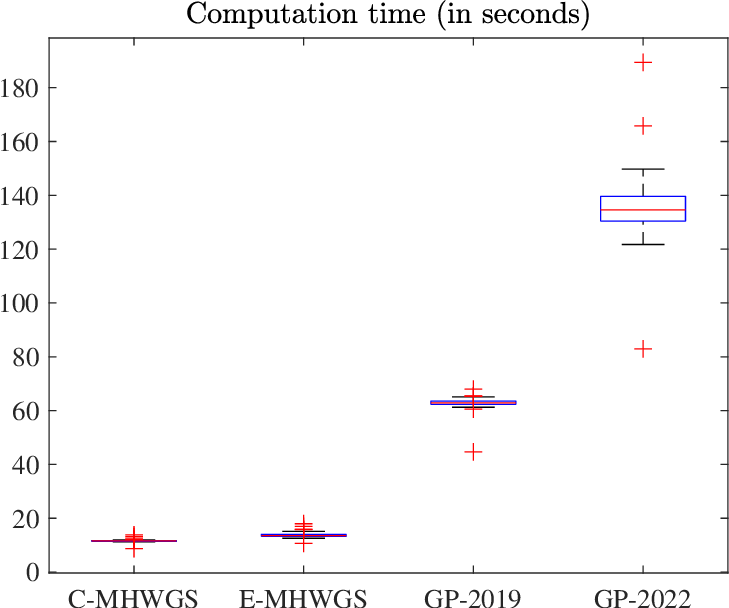}
    \caption{Box plots of the computation time obtained with C-MHWGS, E-MHWGS, GP-2019 and GP-2022 with the 100 Monte Carlo simulations.}
    \label{fig:comp_time_boxplot}
\end{figure}

 The fit for $w$ obtained with the 100 Monte Carlo simulations are represented in box plots in Figure~\ref{fig:best_fit_boxplot} and the computation time in Figure~\ref{fig:comp_time_boxplot}. C-MHWGS gives the worst fit performances with the largest variance and outliers, implying that it did not convergence close to the global optimum most of the times. E-MHWGS is the method with the least outliers and both GP-2019 and GP-2022 give fit very close to 100\% most of the times. While GP-2019 and GP-2022 seem to be the best methods for fitting performances, they however require much more computational power while C-MHWGS is the fastest, preceded closely by E-MHWGS.

\begin{figure*}
    \centering
    \includegraphics[width=\linewidth]{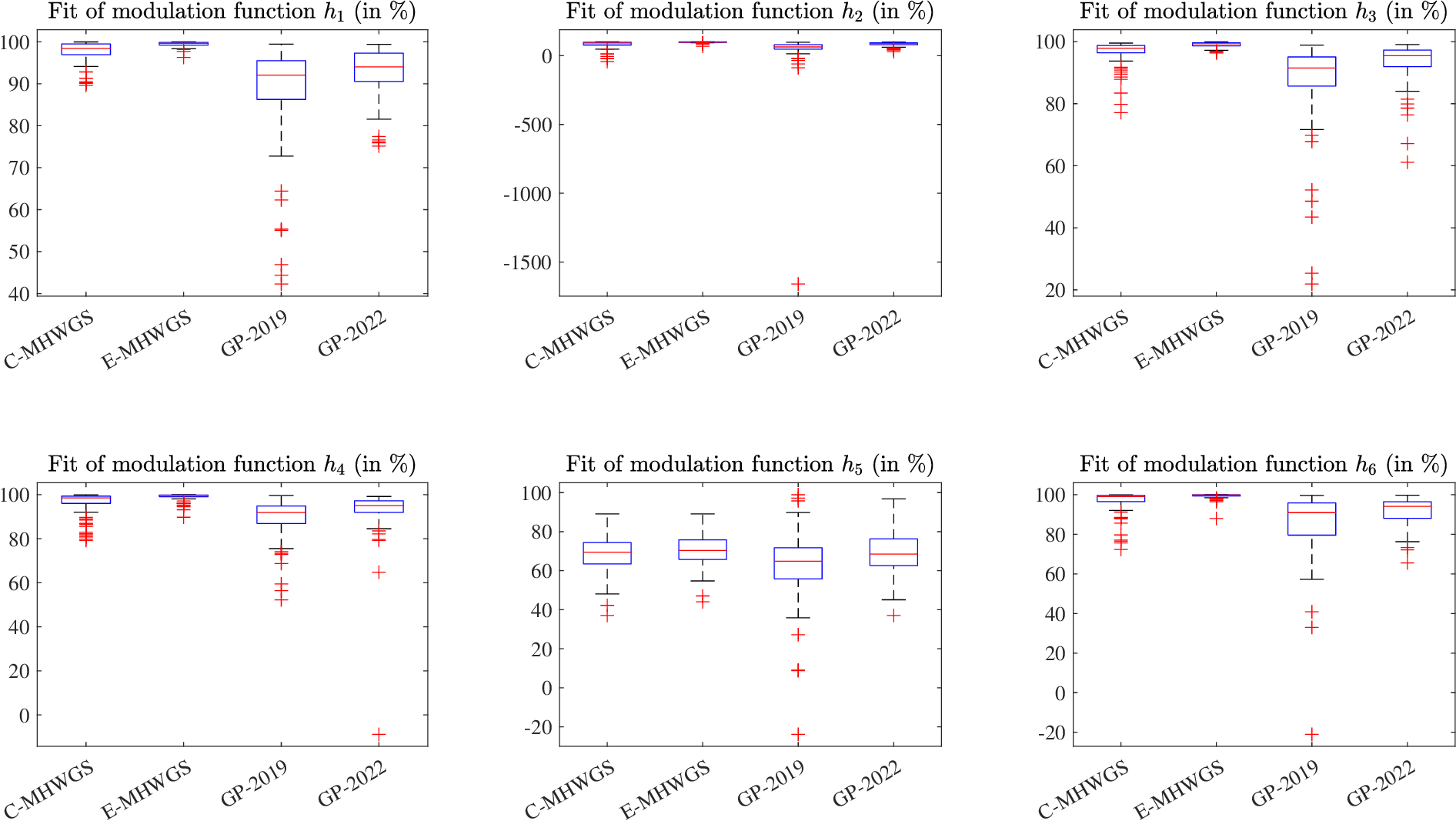}\vspace{0.7cm}
    \includegraphics[width=\linewidth]{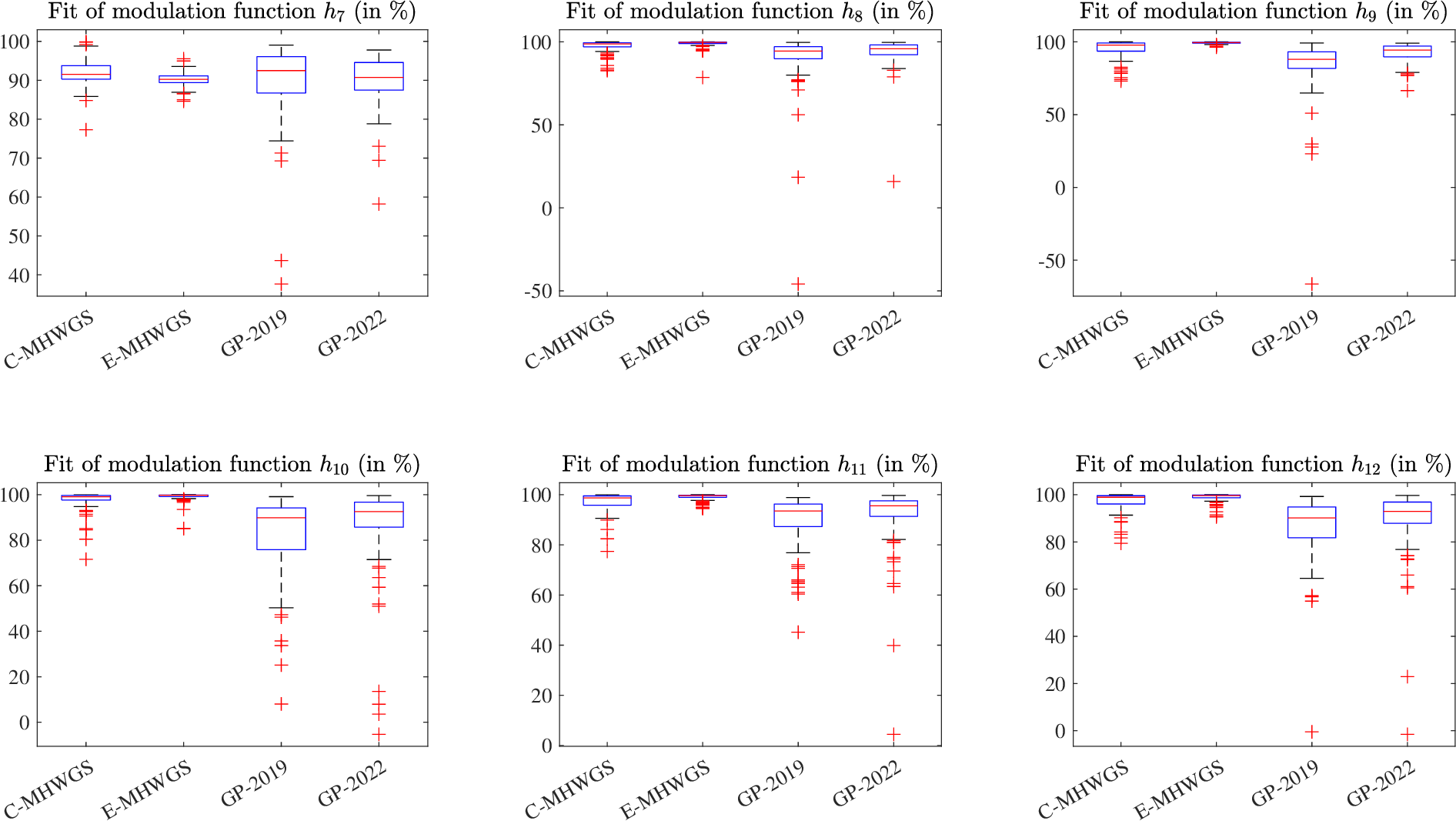}
    \caption{Box plots of the fit of all the modulation functions obtained with C-MHWGS, E-MHWGS, GP-2019 and GP-2022 with the 100 Monte Carlo simulations.}
    \label{fig:fit_h_1}
\end{figure*}

 In Figure~\ref{fig:fit_h_1}, we depict the box plots of the fit for the twelve modulation functions $h_i$. E-MHWGS give the best fit performances for almost all the modulation functions and can be well observed for, e.g.,  $h_1$, $h_2$, $h_9$ and $h_{10}$. C-MHWGS is the second best method and we observe that there is larger variance  outliers for the fitting performances than with E-MHWGS. This can be well observed, e.g., with $h_1$ and $h_3$. The Gaussian process methods give  the worst fitting performances for the modulation function, suggesting that some overfitting occurred. GP-2022 seems to be however slightly better than GP-2019 wince it improves on the fitting performances for all the modulation functions. This was also observed in the numerical example in~\cite{Col:22} with six modulation functions. We also observe large outliers with GP-2019 (for, e.g.,  $h_2$, $h_5$ and $h_9$) and with GP-2019 (for, e.g.,  $h_{4}$, $h_9$ and $h_{12}$). 

\begin{figure}
    \centering
    \includegraphics[width=1\linewidth]{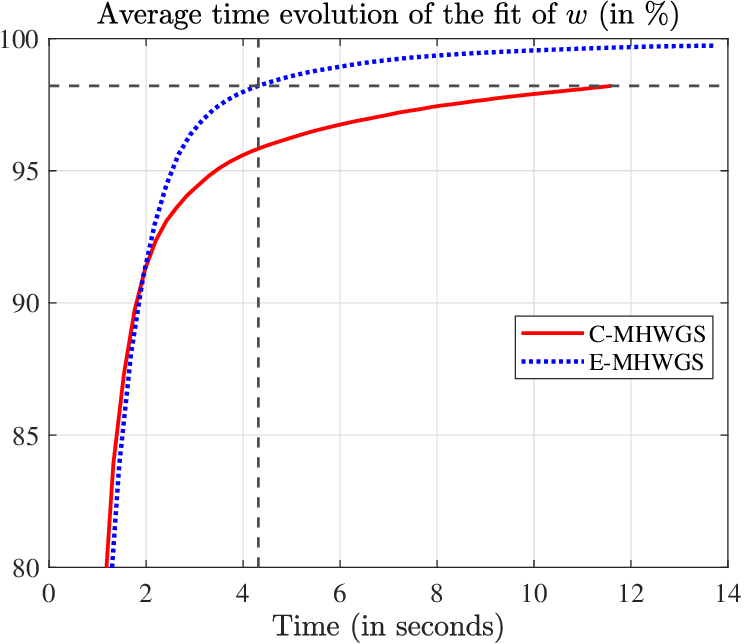}
    \caption{Average time evolution of the fit of $w$ for C-MHWGS (red solid line) and E-MHWGS (blue dotted line). The intersection point of the black vertical and horizontal dashed lines gives the average time it takes E-MHWGS to reach the best average fitting performances obtained with C-MHWGS (at the last EM iteration).}
    \label{fig:evol_fit_time}
\end{figure}

Finally, in Figure~\ref{fig:evol_fit_time}, we plot the average time evolution of the fit on $w$ for E-MHWGS and C-MHWGS and the intersection point which gives the average time it takes E-MHWGS to reach the best average fitting performances obtained with C-MHWGS (at the last EM iteration). We see that E-MHWGS beats C-MHWGS after 4.1s which is less than the half of the time it takes C-MHWGS to achieve its best fitting performances. By observing the last two seconds for both methods, we observe that C-MHWGS could improve with more iterations since its slope is not horizontal while E-MHWGS has a nearly constant fit for the last two seconds. Both aforementioned observations illustrate how faster the proposed sampling scheme is compared to the classical one. 

\section{Conclusion and perspectives}

In this paper, we proposed a Bayesian estimation approach for the nonlinear estimation of Monod kinetics. Using only qualitative prior knowledge about the Monod kinetic parameters, we have chosen a log-Gaussian prior for every kinetic parameter and adapt the Expectation Maximization algorithm for the determination of the log-Gaussian prior  hyperparameters. We have also proposed a new sampling scheme which is a variation of the Metropolis-Hastings within Gibbs sampling, called enforced Metropolis-Hastings within Gibbs sampling, for which the Metropolis-Hastings step is repeated several times in order to increase the change of sample acceptance. It was shown to perform better than the classical Metropolis-Hastings within Gibbs sampling scheme and Gaussian process based methods from the literature on a numerical example.

In future works, we would like to use this estimation technique on real-life data of Chinese Hamster Ovary CHO-K1 cells. We are also interested in analyzing E-MHWGS more theoretically. As an example, we want to verify if the convergence to a stationary probability distribution is still guaranteed as it was the case in, e.g., Gibbs sampling. Finally, other nonlinear estimation problems could be of interest in future studies for which the proposed sampling scheme may be of help.

                             % Place 

\bibliographystyle{plain}        % Include this if you use bibtex 
\bibliography{autosam.bib}           % and a bib file to produce the 
                                 % bibliography (preferred). The
                                 % correct style is generated by
                                 % Elsevier at the time of printing.

%\begin{thebibliography}{99}     % Otherwise use the  
                                 % thebibliography environment.
                                 % Insert the full references here.
                                 % See a recent issue of Automatica 
                                 % for the style.
%  \bibitem[Heritage, 1992]{Heritage:92}
%     (1992) {\it The American Heritage. 
%     Dictionary of the American Language.}
%     Houghton Mifflin Company.
%  \bibitem[Able, 1956]{Abl:56}
%     B.~C.~Able (1956). Nucleic acid content of macroscope. 
%     {\it Nature 2}, 7--9. 
%  \bibitem[Able {\em et al.}, 1954]{AbTaRu:54}   
%     B.~C. Able, R.~A. Tagg, and M.~Rush (1954).
%     Enzyme-catalyzed cellular transanimations.
%     In A.~F.~Round, editor, 
%     {\it Advances in Enzymology Vol. 2} (125--247). 
%     New York, Academic Press.
%  \bibitem[R.~Keohane, 1958]{Keo:58}
%     R.~Keohane (1958).
%     {\it Power and Interdependence: 
%     World Politics in Transition.}
%     Boston, Little, Brown \& Co.
%  \bibitem[Powers, 1985]{Pow:85}
%     T.~Powers (1985).
%     Is there a way out?
%     {\it Harpers, June 1985}, 35--47.

%\end{thebibliography}

\end{document}